\documentclass[%
 reprint,
]{revtex4-2}

\usepackage{amsmath}
\usepackage{graphicx}
\usepackage{dcolumn}
\usepackage{bm}

\begin{document}

\preprint{APS/123-QED}

\title{An attempt to study axion-photon coupling using compact binary \\ systems with high Shapiro time delay}

\author{Zhao-Yang Wang}
\author{Hao-Chen Tian}
\author{Yun-Feng Liang}
    \email{liangyf@gxu.edu.cn}

    \affiliation{Guangxi Key Laboratory for Relativistic Astrophysics, School of Physical Science and Technology, Guangxi University, Nanning 530004, China}

\date{\today}

\begin{abstract}

We study the axion-photon conversion process in pulsar binary systems with high Shapiro time delay.
In these binary systems, the orbital plane is nearly parallel to the line of sight.
When the companion star is positioned between the pulsar and the Earth, the pulsar radiation beam will pass through the companion's magnetic field, potentially leading to axion-photon conversion that affects the intensity and linear polarization of the photons.
The advantage of using such systems for axion or axion-like particle (ALP) research is that the intrinsic intensity and polarization state of the source photons can be well determined.
This work analyzes the axion-photon conversion and the magnetic field along the photon path, establishing the corresponding model and solving it numerically.
We choose PSR J1614-2230 and PSR J1910-5959A for our study. By assuming the companion's magnetic field, magnetic axis angle, and detector sensitivity, we discuss the feasibility of using high Shapiro delay binary systems to constrain the axion/ALP parameter space.
We find that the upper limits derived from these two sources are weaker then existing constraints. 
However, considering the higher reliability, the method proposed in this work is still valuable.
As more suitable samples and richer observational results are discovered, further constraints on the axion parameters could be strengthened.

\end{abstract}

\maketitle


\section{Introduction}
\label{sec1}

Axions and axion-like particles (ALPs) are lightweight pseudoscalar particles that appear in various extensions of the standard model~\cite{Wilczek1978}.
They originally stemmed from the PQ theory proposed by Roberto Peccei and Helen Quinn to explain the strong CP problem in QCD~\cite{Peccei1977_1, Peccei1977_2}.
Additionally, axions are also widely present in string theory~\cite{PeterSvrcek2006}, for example, during the compactification process in string theory, a spectrum of ultralight axions can emerge~\cite{Arvanitaki2010}.
Currently, axions and ALPs are considered among the main candidates for dark matter, as they can be produced non-thermally in the early universe, forming a significant portion of cold dark matter~\cite{axiondarkmatter2023, Preskill1983, Abbott1983, Marsh2011, Chao2024}.
Today, the detection of axions primarily relies on the gravitational effects produced by axions or their interactions~\cite{Baryakhtar2021} with particles in the standard model, using results from laboratory experiments~\cite{Ehret2010, Bahre2013, CASTCollaboration2024, Anastassopoulos2017, Salemi2021, Pandey2024, Gramolin2021, Kahn2016, Braine2020, Braine2021} or astronomical observations~\cite{Gill2011, Dessert2022_1, Dessert2022_2, Dessert2019, Ajello2016, Davies2023, Cheng2021, Abe2024, Noordhuis2023, Jin-Wei2021, Xiao-Jun2021, Hai-Jun2021, Hai-Jun2022, Hai-Jun2023, Hai-Jun2024_1, Hai-Jun2024_2, Lin-Qing2024} to constrain the parameter space of axions and ALPs.

The interaction between axion and the Standard Model is very weak, making them difficult to observe.
Currently, most experiments detect axion and constrain the parameter space of axions through axion-photon interactions.
Laboratory experiments to detect axion-photon interactions include ALPS~\cite{Ehret2010}, CAST~\cite{Anastassopoulos2017, CASTCollaboration2024} and ADMX~\cite{Braine2020, Braine2021}.
Because compact objects and extreme astronomical environments provide natural experimental settings, astrophysical experiments have achieved better detection results.
For example, by studying possible changes in intensity and polarization of the emissions of magnetic white dwarfs~(MWD) and neutron stars~(NS) due to the conversion of photons into axions in strong magnetic fields, or by studying axions converting into X-rays in the magnetic field of a white dwarf~(WD), constraints on the axion parameters have been placed~\cite{Gill2011, Dessert2022_1, Dessert2022_2, Dessert2019}.
Search for spectral irregularities caused by axion-photon oscillations in the $\gamma$ ray spectra of galaxies at the center of clusters~\cite{Ajello2016, Davies2023, Cheng2021, Hai-Jun2023, Hai-Jun2024_1, Hai-Jun2024_2, Lin-Qing2024}.
In the local region of the NS magnetosphere, axions can be produced in large numbers when the surrounding plasma cannot effectively shield the induced electric field, they can be resonantly converted into photons that can be detected~\cite{Noordhuis2023}.

In this work, we will select pulsar binary systems with high Shapiro delay for axion-like particle research.
The Shapiro delay, also known as the gravitational time delay of light, refers to the delay effect of light passing through a strong gravitational field.
Through the Shapiro delay, we can precisely infer the masses of the pulsar and its companion, as well as the orbital inclination angle \cite{Jacoby2005, Possel2020, Verbiest2008}.
In pulsar binary systems with high Shapiro delay, the orbital inclination angle is close to $\pi/2$.
The pulsar signal may undergo axion-photon oscillations as it passes through the companion's magnetic field, leading to variations in light intensity and linear polarization.
Due to the high stability of pulsar signals, these changes can be easily identified and resolved, allowing for constraints on the axion parameter space with low uncertainty.
In Sec.~\ref{sec3}, we select two pulsar binary systems with high Shapiro delay for specific analysis, discuss the results of the constraint on the axion parameter space under different circumstances, and hope to observe more suitable pulsar binary systems in the future.

The structure of this paper is arranged as follows.
In Sec.~\ref{sec2}, we start from the axion Lagrangian to construct the Euler-Lagrange equations to study the effects of axion-photon oscillations and analyze the magnetic field distribution along the line of sight.
In Sec.~\ref{sec3}, we specifically analyze the variations in light intensity and linear polarization caused by axions in different binary systems.
In Sec.~\ref{sec4}, we discuss the feasibility of constraining the axion parameter space using pulsar binary systems with high Shapiro delay and provide suggestions for future observations.

\section{Model Equations}
\label{sec2}

    \subsection{Axion-Photon Mixing Equations}
In this section, we start by deriving the equations of motion, provide a brief overview of axion-photon coupling theory, and discuss its effects on the intensity and polarization of light sources in astrophysical environments.
Interactions between the axion and photon are described by the following Lagrangian
\begin{equation}
\begin{aligned}
    \mathcal{L} = & \frac12 (\partial_\mu a \partial^\mu a - m_a^2 a^2)-\frac14 F_{\mu\nu}F^{\mu\nu}\\
    & -\frac{g_{a\gamma\gamma}}{4}aF_{\mu\nu}\widetilde{F}^{\mu\nu}\\
    & +\frac{\alpha_{\rm em}^2}{90m_e^4}\left[(F_{\mu\nu}F^{\mu\nu})^2+\frac74(F_{\mu\nu}\widetilde{F}^{\mu\nu})^2 \right]
\label{Lagrangian density}
\end{aligned}
\end{equation}
where $a$ is the axion field, $m_a$ is the axion mass, $g_{a\gamma\gamma}$ is the coupling constant between axion and photon, $\alpha_{\rm em}$ is the fine-structure constant of electromagnetism, $m_e$ is the electron mass, and $F^{\mu\nu}$ is the electromagnetic field tensor.
The first term represents the axion field, the second term represents the photon field, the third term is the axion-photon coupling, and the fourth term known as the Euler-Heisenberg Lagrangian, which describes the self-interaction of photons below the electron mass, specifically accounting for the effects of QED vacuum birefringence in strong magnetic fields.

First, we discuss the axion-photon coupling in a uniform magnetic field.
We establish a Cartesian coordinate system where the uniform magnetic field ${\mathbf{B}}=B_0\hat{e_x}$ is parallel to the $x$-axis, and photons are incident from the $z$-axis, traveling through a distance of length $L$.
By substituting the Lagrangian~(\ref{Lagrangian density}) into the Euler-Lagrange equation, we obtain the equation of motion for the field.
Assuming that the magnetic field range is much larger than the photon wavelength ($L \gg 2\pi/\omega$) and that the axion mass is small, the axion velocity is approximately the speed of light.
The equation of motion is changed into a system of first order differential equations by WKB approximation\cite{Raffelt1988}.
\begin{equation}
\begin{aligned}
    \left[ i\partial_z +
    \left(
    \begin{array}{ccc}
        \Delta_\perp & 0 & 0 \\
        0 & \Delta_\parallel & \Delta_B \\
        0 & \Delta_B & \Delta_a
    \end{array}
    \right) \right]
    \left(
    \begin{array}{c}
        A_\perp \\
        A_\parallel \\
        a
    \end{array}
    \right) = 0
    \label{axion-photon mixing equations}
\end{aligned}
\end{equation}
where
\begin{equation}
\begin{aligned}
    \Delta_\perp & = 2 \omega \xi &&
    & \Delta_\parallel & = \frac72 \omega \xi \\
    \Delta_B & = \frac12 g_{a\gamma\gamma} B_0 &&
    & \Delta_a & = -\frac{m_a^2}{2 \omega}
    \label{axion-photon mixing parameter 1}
\end{aligned}
\end{equation}
In Eq.~(\ref{axion-photon mixing equations}), $A_\parallel$ and $A_\perp$ are the components of the photon along the $x$-axis and $y$-axis, respectively, and $a$ is the axion field.
$\Delta_\perp$ and $\Delta_\parallel$ represent the Euler-Heisenberg effect in QED, where $\xi = (\alpha_{\rm em}/45\pi )(B_0/B_{\rm crit})^2$ and the critical magnetic field strength $B_{\rm crit}=m_e^2/e \approx 4.41\times 10^{13}\,\rm{G}$~\cite{Heisenberg2006}.
Generally, $\Delta_\parallel$ also includes plasma terms $\Delta_\omega = -\omega_{pl}^2/2\omega$, with plasma frequency $\omega_{pl}$ depending on the electron density, the free electron density in the interstellar medium may be as much as $n_e \sim 10^{-1}/ \rm{cm^3}$~\cite{Cordes2002}, and the effect of the plasma term is negligible in this work.
$\Delta_B$ represents the mixing term, describing the conversion between photons and axions in the background magnetic field. 
$\Delta_a$ includes the axion mass and photon frequency, representing the difference in momentum between axions and photons.

Solving this equation demonstrates that axion and photon convert into each other periodically, which leads to the phenomenon known as axion-photon oscillation.
From Eq.~(\ref{axion-photon mixing equations}), it is evident that only photons and axions parallel to the magnetic field direction possess a mixing term $\Delta_B$, indicating that only photons in this direction will participate in the axion-photon oscillation.
This means that axion-photon oscillations not only affect light intensity but also influence polarization, and the presence of axions can be detected indirectly through electromagnetic waves passing through astronomical environments with strong magnetic fields.

Since the direction of the magnetic field in the actual astronomical environment will not remain in the same plane, the above equation is not suitable for practical work, and it should be transformed into a coordinate system independent of the magnetic field, which can be achieved through a simple coordinate transformation.
\begin{equation}
\begin{aligned}
    \left[ i\partial_z +
    \left(
    \begin{array}{ccc}
        \Delta_{xx} & \Delta_{xy} & \Delta_{B_x} \\
        \Delta_{xy} & \Delta_{yy} & \Delta_{B_y} \\
        \Delta_{B_x} & \Delta_{B_y} & \Delta_a
    \end{array}
    \right) \right]
    \left(
    \begin{array}{c}
        A_x \\
        A_y \\
        a
    \end{array}
    \right) = 0
    \label{General axion-photon mixing equations}
\end{aligned}
\end{equation}

Generally, the magnetic field distribution is a function of position, and the transformation process between axions and photons can be obtained by substituting the specific magnetic field function into Eq.~(\ref{General axion-photon mixing equations}) and solving it.
Various hybrid terms in Eq.~(\ref{General axion-photon mixing equations}) are defined as follows in Eqs.~(\ref{axion-photon mixing parameter 2}), with $\xi_x = (\alpha_{\rm em}/45\pi )(B_x/B_{\rm crit})^2$ and $\xi_y = (\alpha_{\rm em}/45\pi )(B_y/B_{\rm crit})^2$.
\begin{equation}
\begin{aligned}
    \Delta_{B_x} & = \frac12 g_{a\gamma\gamma} B_x &&
    & \Delta_{B_y} & = \frac12 g_{a\gamma\gamma} B_y \\
    \Delta_{xx} & = 2 \omega (\frac74 \xi_x + \xi_y) &&
    & \Delta_{yy} & = 2 \omega (\xi_x + \frac74 \xi_y) \\
    \Delta_{xy} & = \frac{3 \alpha_{\rm em} \omega B_x B_y}{90 \pi B_{\rm crit}^2} &&
    & \Delta_a & = -\frac{m_a^2}{2 \omega}
    \label{axion-photon mixing parameter 2}
\end{aligned}
\end{equation}

We are concerned with the conversion probability from photons to axions, as this can be verified by observing light intensity and linear polarization.
The conversion probability and linear polarization are expressed by Eq.~(\ref{polarization}), with the linear polarization of photons represented by Stokes parameters.
\begin{equation}
\begin{aligned}
    p_{\gamma \to a} & = \frac{|a|^2}{|A_x|^2+|A_y|^2+|a|^2} \\
    L_p & = \frac{\sqrt{Q^2+U^2}}{I}
    \label{polarization}
\end{aligned}
\end{equation}
where $I=|A_x|^2+|A_y|^2$ represents the radiation intensity, $Q=|A_x|^2-|A_y|^2$ and $U=2Re(A_x A_y^*)$ parameterize the linear polarization.

    \subsection{Binary Magnetic Field Models}
For a pulsar binary system with a highly inclined orbital plane, the pulsar will experience nearly coincident periods with the companion star during each orbital cycle.
In such cases, the Shapiro delay can be used to infer the companion star's mass and the orbital inclination.
Shapiro delay refers to the additional time light takes to travel through the curved spacetime near a massive object.
This subtle delay can be observed using millisecond pulsars or pulsars with very stable pulse radiation periods.

\begin{figure}[htbp]
    \centering
    \includegraphics[width=1\linewidth]{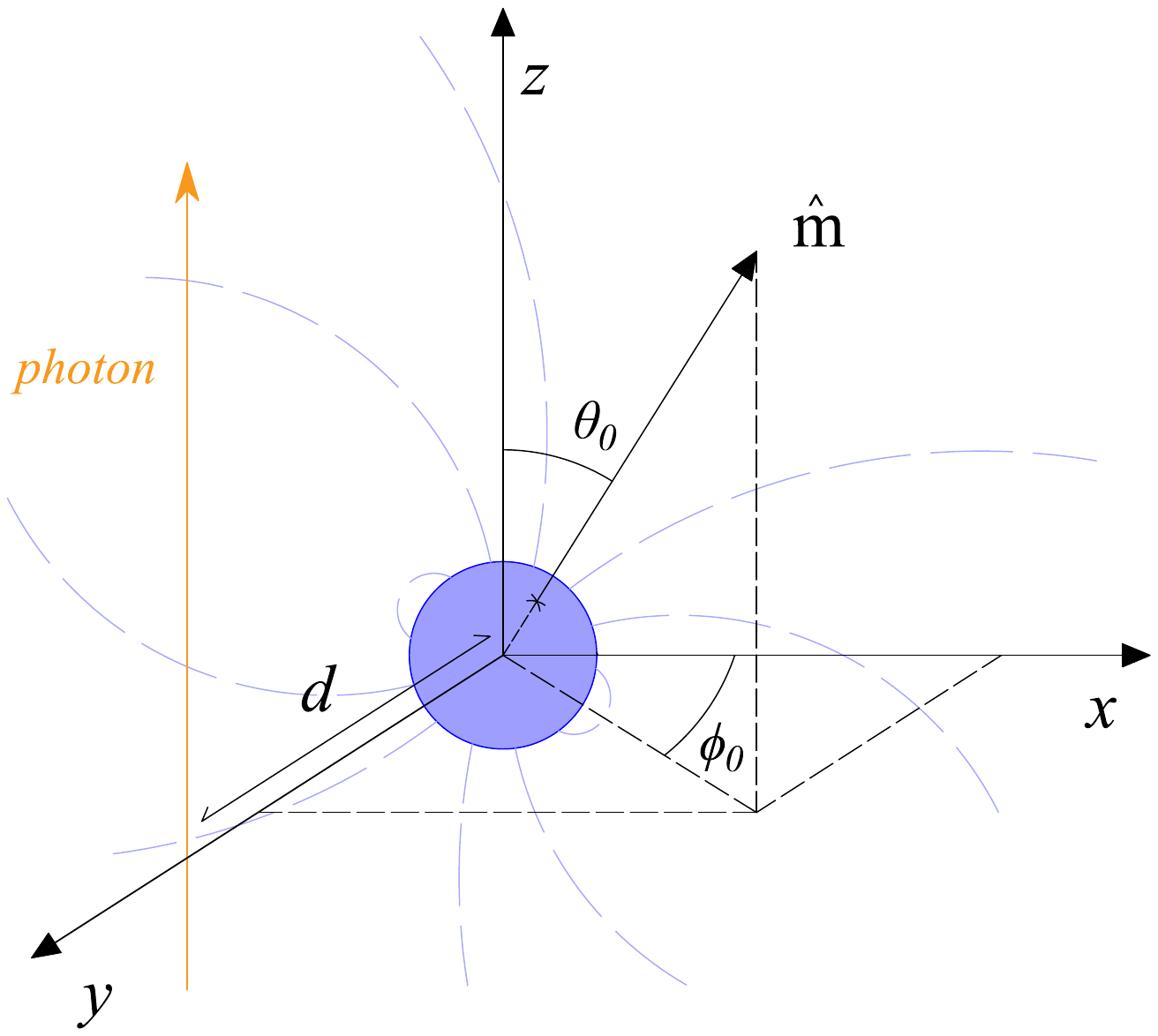}
    \caption{\label{coordinate system} The coordinate system used in this work. The origin is at the center of the WD. The dipole magnetic field unit vector is $\hat{\bf{m}}$.  
    The pulsar signal intersects the $y$-axis and propagates along the $z$-axis. $d$ is the perpendicular distance between the photon path (orange line) and the WD center.}
\end{figure}

In this section, we will discuss the magnetic field distribution along the pulsar signal path in pulsar binary systems with significant Shapiro delay.
The magnetic field distribution around a compact object approximates a dipole structure, and WD magnetic field can be modeled by superimposing the dipole component with a multipole component, where the contribution of the multipole field decreases more rapidly as the WD radius increases.
In this work, since the pulsar signal is at a considerable distance (greater than one order of magnitude) from the WD, the contribution of the multipole field can be ignored, and only the dipole magnetic field distribution will be considered.
\begin{equation}
\begin{aligned}
    {\mathbf{B}}({\mathbf{r}},{\mathbf{m}}) = \frac{B_s}{2} \left( \frac{r_0}{r} \right)^3 [3(\hat{\mathbf{m}} \cdot \hat{\mathbf{r}})\hat{\mathbf{r}} - \hat{\mathbf{m}}]
    \label{dipole form}
\end{aligned}
\end{equation}
where $r_0$ is the radius of the WD, $B_s$ is the surface magnetic field strength of the WD, $\hat{\bf{r}}$ represents the radial unit vector, and $\hat{\bf{m}}$ represents the dipole unit vector — the direction of the magnetic axis.
It can be seen that the magnetic field depends only on the position and the magnetic axis.
The coordinate system is centered at the WD, as shown in Fig.~\ref{coordinate system}, where $\hat{\bf{m}}$ is as follows.
\begin{equation}
\begin{aligned}
    \hat{\bf{m}} = & \sin(\theta_0)\cos(\phi_0)\hat{\bf{x}} + \sin(\theta_0)\sin(\phi_0)\hat{\bf{y}} + \cos(\theta_0)\hat{\bf{z}}
    \label{angle}
\end{aligned}
\end{equation}

In general, the spin period of WD is much longer than the time required for a pulsar signal to pass through its vicinity.
Therefore, the shift in the magnetic axis direction caused by the WD's spin is negligible, and the spin of the WD is not considered in the calculation of the axion-photon propagation equation.
By varying $\theta_0$ and $\phi_0$, any angle of the magnetic axis can be considered.
It is worth mentioning that when the WD approaches or recedes from the line of sight, changes in the magnetic field distribution cause the pulsar signal to exhibit a brightness variation lasting tens of minutes to about an hour within one orbital period of the binary system. 

\section{Discussion of constrained axion parameter space}
\label{sec3}

In this section, we use the previously discussed analysis of axion-photon oscillation and path magnetic fields to solve the axion-photon propagation equation.
We then apply this to compact binary systems that meet the criteria and discuss the constraints on axion mass, ALP mass, and axion-photon coupling parameters through numerical solution equations.
The coordinate system used to solve the equations is illustrated in Fig.~\ref{coordinate system}, where photon propagation occurs along the $y$-axis, with its direction perpendicular to the $z$-axis, and there is a certain distance $d$ from the WD, which is obtained by measuring the Shapiro delay.
In our study, both the initial and final positions of the equations are situated five solar radii from the WD.
At these two points, we find that magnetic field intensity $B \sim 1\,\rm{G}$, the axion-photon conversion can be considered negligible.
Our focus is on the region closer to the WD, where the magnetic field is stronger.
Here we also ignore the influence of gravitational microlensing, because the deflection angle of the light caused by the gravitational field of the companion star is very small ($\theta \sim 10^{-5}$), and the deviation of the light path is negligible.

In the axion-photon propagation equation, the axion conversion probability depends on the magnetic field strength perpendicular to the propagation direction, with higher magnetic field strength leading to greater conversion efficiency.
We select two NS-WD binary systems with high Shapiro delays as reference sources: PSR J1614-2230 and PSR J1910-5959A.

    \subsection{PSR J1614-2230}
PSR J1614-2230 is an NS-WD system, and Demorest et al.~\cite{Demorest2010} were the first to measure the Shapiro delay in this system.  
According to the NANOGrav 15-Year Data Set~\cite{NANOGrav2023}, the pulsar mass is $M_{p} = 1.937 \pm 0.014\,M_{\odot}$, and the WD companion star mass is $M_{c} = 0.494 \pm 0.002\,M_{\odot}$.
Shamohammadi et al.~\cite{Shamohammadi2023} measured the angle between the rotation axis of the binary system and the line of sight as $i = 89.179^{\circ} \pm 0.013^{\circ}$, and the closest distance at which the pulsar's radiation passes through the WD is $d \sim 2.4 \times 10^{8}\,\rm{m}$.
For this WD companion, considering the electron degeneracy pressure provides a rough estimate of the WD radius $R_{\rm wd} \sim 9.87 \times 10^{6}\,\rm{m}$, which is in close agreement with the empirical formula given by Nauenberg et al.~\cite{Nauenberg1972}.
Next, we discuss the magnetic field of WD, for low or zero magnetic field WD, the magnetic field along the pulsar's propagation path is too weak to produce a sufficiently large conversion probability due to axion-photon oscillations.
Here, we assume the companion star is a MWD, based on the study of MWD by Ferrario et al.~\cite{Ferrario2015}, we selected three typical surface magnetic field strengths for this MWD ($\rm{1\,MG}$, $\rm{10\,MG}$, $\rm{100\,MG}$).

\begin{figure*}[htbp]
    \centering
    \includegraphics[width=0.9\linewidth]{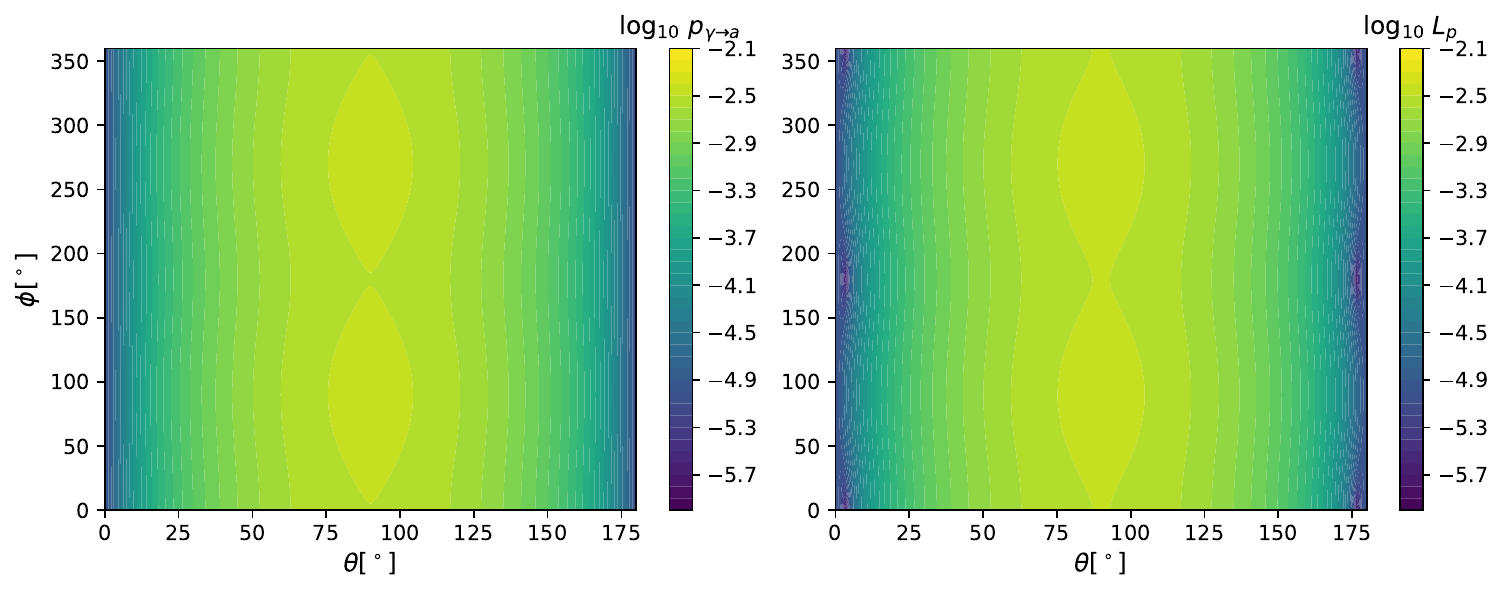}
    \caption{\label{L&P_theta-phi} The photon-to-axion conversion probability (left panel) and linear polarization degree (right panel) for different magnetic axis angles as pulsar photons pass by the companion WD in the PSR J1614-2230 binary system.
    The axion-photon coupling constant $g_{a\gamma\gamma}=1 \times 10^{-9}\,\rm{GeV^{-1}}$ and the axion mass $m_a = 1 \times 10^{-8}\,\rm{eV}$ are adopted for illustration.
    The color represents the magnitude of the axion conversion probability.
    When $\theta = 0$ ($\theta = \pi/2$ and $\phi = \pi/2$), the magnetic axis is parallel (perpendicular) to the line of sight direction, resulting in the lowest (highest) conversion efficiency.
    Under the assumption of initially unpolarized pulsar photons, the trend in linear polarization degree is highly correlated with the conversion probability.}
\end{figure*}

By substituating the parameters of PSR J1614-2230 into the Eq.~(\ref{General axion-photon mixing equations}), we can get the photon-to-axion conversion probability and the photon polarization degree under different conditions.
Here, we first calculate using natural light with an energy of $1\rm{eV}$, setting the axion-photon coupling constant to $g_{a\gamma\gamma}=1 \times 10^{-9} \rm{GeV^{-1}}$ and the axion mass to $m_a = 1 \times 10^{-8} \rm{eV}$.
We calculate the conversion probabilities and polarization degrees for various magnetic axis angles, as shown in Fig.~\ref{L&P_theta-phi}.
The left panel shows the variation in axion conversion probability, while the right panel shows the variation in photon linear polarization degree, with a high correlation between the two.
We find that the axion conversion probability and photon polarization degree are minimized when the magnetic axis is parallel to the line of sight direction.
This occurs because on the path of photon propagation, $B_y = 0$, only $B_x$ exists, and $B_x$ centered at $z=0$ is equal in magnitude and opposite in direction.
When the photons is brought close to the WD, some of the photons are first converted into axions under the influence of the magnetic field, and when the photons is moved away from the WD, this part of the axions will be changed back into photons under the influence of the reverse magnetic field.
The axion conversion probability and photon polarization degree are maximized when the magnetic axis is perpendicular to the line of sight direction, that is, when $\theta_0 = \pi/2$ and $\phi_0 = \pi/2$.
At this orientation, $B_y$ along the propagation path is relatively large, which promotes the conversion of photons into axions, with the axion conversion probability reaching $p_{\gamma \to a}=0.347\%$.
In a real astronomical environment, the orbital motion of the binary system and the rotation of the companion star must be considered.
During each orbital period, the angle between the magnetic axis of the WD and the photon path changes.
Except for the extreme case where the magnetic axis is parallel to the propagation direction, any other angle will produces variations in both light intensity and linear polarization degree.
By comparing the signals observed during the rest of the orbital period, the effects caused by axions can be distinguished.

\begin{figure}[htbp]
    \centering
    \includegraphics[width=1\linewidth]{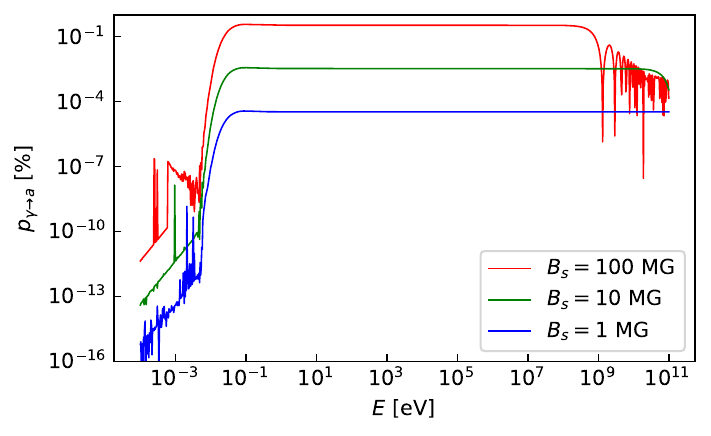}
    \caption{\label{P_v_B} In the PSR J1614-2230 binary system, with the magnetic axis angle of the WD set to $\theta_0 = \pi/2$ and $\phi_0 = \pi/2$, the axion-photon coupling $g_{a\gamma\gamma}=1 \times 10^{-9}\,\rm{GeV^{-1}}$, and the axion mass $m_a = 1 \times 10^{-8}\,\rm{eV}$, we calculate the axion conversion probability of pulsar beam passing through the WD surrounding for different magnetic field assumptions.
    The red, green, and blue lines represent the axion conversion probabilities for magnetic field strengths of $1\,\rm{MG}$, $10\,\rm{MG}$, and $100\,\rm{MG}$ on the surface of the WD, respectively.}
\end{figure}

In Fig.~\ref{L&P_theta-phi}, we obtain the magnetic axis angles for maximum conversion efficiency, $\theta = \pi/2$ and $\phi = \pi/2$, and then substitute these into the subsequent calculations to discuss the effects of different WD magnetic field strengths on axion conversion probability and photon linear polarization degree.
We select photons of various energies and numerically solve the axion-photon oscillation equations for different magnetic field strengths.
Since natural light is used as the initial condition, the results for axion conversion probability and photon linear polarization degree are similar, so only the axion conversion probability is shown.
As depicted in Fig.~\ref{P_v_B}, the magnetic field strength has a significant impact on both conversion probability and linear polarization degree, approximately following $p_{\gamma \to a} \propto B_s^2$.
This is consistent with the results obtained by Dessert et al.~\cite{Dessert2022_1} using the weak-mixing approximation.
When the photon energy is $E \lesssim 0.1 \rm{eV}$ or $E \gtrsim 10^9 \rm{eV}$, the conversion probability under different magnetic fields decreases rapidly, which can be explained here using Eq.~(\ref{axion-photon mixing equations}).
For low photon energies, $\Delta_a \gg \Delta_B$, so a higher magnetic field is needed to enhance axion production.
For high photon energies, $\Delta_\parallel \gg \Delta_{B}$, since $\Delta_\parallel$ is proportional to the square of the magnetic field strength, while $\Delta_{B}$ is only proportional to the magnetic field strength, meaning a very high magnetic field suppresses axion production at the same photon energy.
In the central, stable region of Fig.~\ref{P_v_B}, which represents the strong-mixing region, the photon energy ranges from $0.1 \rm{eV} \lesssim E \lesssim 10^9 \rm{eV}$, and the conversion probabilities for photons of different energies are similar.
It is noteworthy that near $0.1 \rm{eV}$, there is a peak in the axion conversion probability, which is related to the magnetic field structure and the axion mass.
Observing the corresponding wavelength ranges can provide better constraints on the axion parameter space.

\begin{figure}[htbp]
    \centering
    \includegraphics[width=1\linewidth]{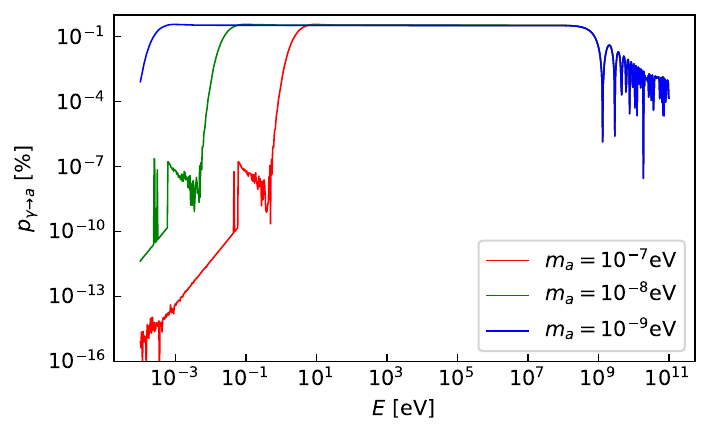}
    \caption{\label{P_v_ma} 
    Similar to Fig.~\ref{P_v_B} but choosing the surface magnetic field strength to be $B=100\,\rm{MG}$ and considering different axion masses of $10^{-7}\,\rm{eV}$, $10^{-8}\,\rm{eV}$ and $10^{-9}\,\rm{eV}$.
    }
\end{figure}

We now consider the effects of different axion masses on the axion conversion probability and photon polarization degree.
With WD surface magnetic field strength $B=100\rm{MG}$, the results for the axion conversion probability are shown in Fig.~\ref{P_v_ma}.
Three axion masses are selected as examples ($10^{-7} \rm{eV}$, $10^{-8} \rm{eV}$, $10^{-9} \rm{eV}$), with the photon energy $E$ as the independent variable, and changing the axion mass results in a shift of the conversion probability curve.
Since $\Delta_a \propto m_a^2$, an increase in axion mass makes it easier for $\Delta_a$ to be larger than $\Delta_B$, which requires higher photon energies to maintain a high conversion probability, and also results in a reduction of the strong mixing region.
Along with the shift of the axion conversion probability curve, the locations of the extrema change, demonstrating that photons of different energies have different sensitivities to varying axion mass.

\begin{figure}[htbp]
    \centering
    \includegraphics[width=1\linewidth]{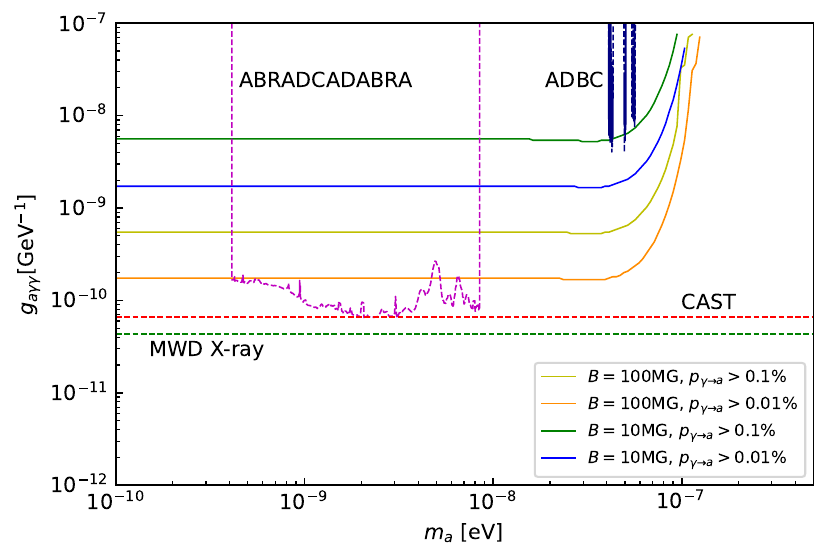}
    \caption{\label{garr_ma_j1614-2230} Expected limits on axion parameter space that could be derived from the optical (energy of $1\,\rm{eV}$ for this plot) observation of PSR J1614-2230, with the assumption of the WD magnetic axis angles of $\theta_0 = \pi/2$ and $\phi_0 = \pi/2$ and surface magnetic field strengths of $B=100\,\rm{MG}$ or $B=10\,\rm{MG}$. 
    The dashed lines in the figure are from other experiments~(Ref.~\cite{Axionlimits2020, Anastassopoulos2017, CASTCollaboration2024, Salemi2021, Pandey2024, Dessert2022_2}). 
    {The $p_{\gamma \to a}>0.1\%$ and $p_{\gamma \to a}>0.01\%$ assume that the observations can reach $0.1\%$ and $0.01\%$ precision, respectively (i.e., have the ability to identify intensity variation of $10^{-3}$ and $10^{-4}$).}
    The strongest limits can exclude the axion-photon coupling parameter $g_{a\gamma\gamma}>1.75\,\times 10^{-10} \rm{GeV^{-1}}$ for the axion mass $m_a<4.10 \times 10^{-8}\, \rm{eV}$.}
\end{figure}

To obtain the upper limit on the axion parameter space, we compute the axion conversion probabilities for different axion-photon coupling parameters and axion masses, with the magnetic axis angles set to $\theta_0 = \pi/2$ and $\phi_0 = \pi/2$.
Assuming the detector precision is on the order of one thousandth or one ten-thousandth, we consider that when the change of light intensity is less than the detector’s accuracy, the generation of axions cannot be observed.
This is a rough estimate, here, we only conduct a feasibility analysis and discuss whether this method can better restrict the axion parameter space.
The paper shows the limiting results obtained by detecting natural light with an energy of $1 \rm{eV}$, where the WD surface magnetic field strengths are $10 \rm{MG}$ and $100 \rm{MG}$ respectively, as shown in Fig.~\ref{garr_ma_j1614-2230}.
The red dashed line represents the constraint from CAST~\cite{Anastassopoulos2017, CASTCollaboration2024}, the purple dashed line represents the constraint from ABRADABRA Demonstrator~\cite{Salemi2021}, the blue dashed line represents the constraint from ADBC~\cite{Pandey2024}, and the green dashed line represents the constraint from MWD X-ray~\cite{Dessert2022_2}.
Plotting scripts and limit data are available at Ref.~\cite{Axionlimits2020}.
The solid line indicates the results of this work.
As previously analyzed, when the axion mass is large, the $1 \rm{eV}$ photon exits the strong mixing region, the conversion efficiency becomes smaller and the limiting result is weak, while limiting result is better near $m_a \sim 10^{-7} \rm{eV}$.
{Taking $B_s=100\,\rm{MG}$ and $p_{\gamma \to a}>0.01\%$~(orange) as an example, we constrain the axion-photon coupling parameter to $g_{a\gamma\gamma}<1.69 \times 10^{-10} \rm{GeV^{-1}}$ for the axion mass $m_a<4.10 \times 10^{-8} \rm{eV}$.}
These results are weaker compared to the current most stringent limits in this region, which are $g_{a\gamma\gamma}=5.4 \times 10^{-12} \rm{GeV^{-1}}$ from Dessert et al.~\cite{Dessert2022_1}.

    \subsection{PSR J1910-5959A}
\begin{figure}[htbp]
    \centering
    \includegraphics[width=1\linewidth]{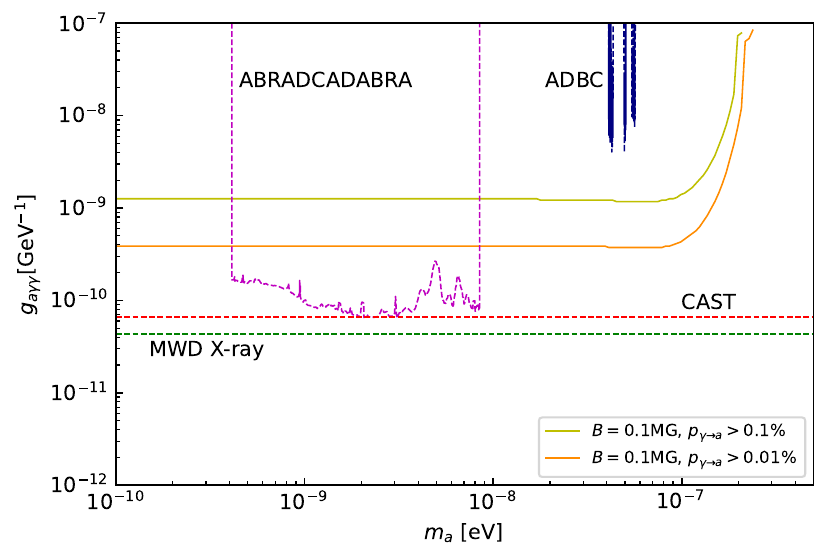}
    \caption{\label{garr_ma_j1910-5959A} 
    Similar to Fig.~\ref{garr_ma_j1614-2230} but for the source PSR J1910-5959A. The parameters are $\theta_0 = \pi/2$, $\phi_0 = \pi/2$ and $B=0.1\,\rm{MG}$.
    The strongest limits can exclude the axion-photon coupling parameter $g_{a\gamma\gamma}>3.87 \times 10^{-10} \rm{GeV^{-1}}$ for the axion mass $m_a<8.21 \times 10^{-8} \rm{eV}$.}
\end{figure}

PSR J1910-5959A is also an NS-WD system, and Corongiu et al.~\cite{Corongiu2023} measured the Shapiro delay in this binary system.
The angle between the rotation axis of the binary system and the line of sight is $i = 88.9^{\circ} \pm 0.14^{\circ}$.
The pulsar mass $M_{p} = 1.55 \pm 0.06M_{\odot}$, and the WD companion star mass $M_{c} = 0.202 \pm 0.006M_{\odot}$, with an orbital period of $P_{B} = 0.8371\,\rm{days}$.
This is a low eccentricity system, and from the concentric orbits it is possible to calculate the closest distance $d \sim 6.02 \times 10^{7} \rm{m}$ when the pulsar radiation passes by the WD.
Bassa et al.~\cite{Bassa2006} analyzed the spectrum of the WD, and according to Istrate et al.~\cite{Istrate2016}, inferred the radius of the WD to be $R_{c} = 0.043 \pm 0.001 R_{\odot}$.
However, no significant zeeman splitting was observed in the spectrum, indicating that the WD companion of PSR J1910-5959A is not a MWD.
According to the optical zeeman spectrum analysis of MWDs by Euchner et al.~\cite{Euchner2002}, the surface magnetic field of the WD is significantly less than $1 \rm{MG}$.
In this work, we assume a magnetic field strength of $B_s=0.1 \rm{MG}$ and substitute it into Eq.~\ref{General axion-photon mixing equations} for numerical solving.
Keeping other parameters consistent with PSR J1614-2230, the axion conversion probability is $p_{\gamma \to a}=0.067\%$, which is worse than the result for PSR J1614-2230.
Based on the analysis in the previous section, the change of axion conversion probability brought about by magnetic field strength and axion mass is not discussed here, and only present the exclusion limits of PSR J1910-5959A on the axion parameter space, as shown in Fig.~\ref{garr_ma_j1910-5959A}.
The dashed lines in the figure represent other experimental data~(Ref.~\cite{Axionlimits2020, Anastassopoulos2017, CASTCollaboration2024, Salemi2021, Pandey2024, Dessert2022_2}), while the solid line indicates the results of this work.
Similar to the results for J1614-2230, for $B_s=0.1 \rm{MG}$ and $p_{\gamma \to a}>0.01\%$~(orange), the axion-photon coupling parameters $g_{a\gamma\gamma}>3.87 \times 10^{-10} \rm{GeV^{-1}}$ and the axion mass in the range of $m_a>8.21 \times 10^{-8} \rm{eV}$ are restricted.

\section{Conclusion}
\label{sec4}

This work investigates the variations in pulsar emission flux and linear polarization degree resulting from axion-photon conversion in pulsar binary systems with significant Shapiro delay. 
We are interested in the time period of an orbital period when the companion is closest to the line of sight, during which the intensity and linear polarization of the pulsar signal will change under the influence of the magnetic field of the companion. 
In contrast, during the rest of the orbital period, the highly stable pulsar signal provides a reference for the true/unmodulated source flux.
The variation in the intensity is represented by the axion conversion probability ($p_{\gamma \to a}$), while the linear polarization degree ($L_p$) is given by the Stokes parameter.
We choose PSR J1614-2230 and PSR J1910-5959A for the discussion, and analyze the influence of different magnetic field configurations on the axion conversion probability and the photon linear polarization degree.
Analyzing pulsar binary systems with significant Shapiro delay represents a new astrophysical method to constrain the axion parameter space, which has lower uncertainty compared to some other astrophysical detection methods because the initial source flux can be reliably determined.

Although we find PSR J1614-2230 and PSR J1910-5959A are already the two most promising systems among all known pulsar binaries for studying the axion-photon coupling, the observational effects due to axions in these two systems are still too small for probing the parameter space of current interest (i.e., $g_{a\gamma\gamma}<10^{-11}\,{\rm GeV^{-1}}$), so this work does not use real data to place constraints but only discuss the possible probe/exclusion capability.
We derive the axion parameter space that could be constrained by the PSR J1614-2230 and PSR J1910-5959A observations.
The expected constraints are better for PSR J1614-2230, as shown in Figs.~\ref{garr_ma_j1614-2230} and \ref{garr_ma_j1910-5959A}.
Assuming a detection precision of 1/1000, the method proposed in this paper has the potential to detect $g_{a\gamma\gamma}\sim5\times10^{-10}$ of the parameter space, which is significantly weaker than the current strongest constraints on the axion parameters ($g_{a\gamma\gamma}\lesssim5\times 10^{-12}\,{\rm GeV^{-1}}$, e.g. \cite{Dessert2022_1,Noordhuis2023}).
However, considering the less uncertainty in determining the unmodulated photon flux, a study of the axion-photon conversion effect in pulsar binary systems would still be valuable.

In addition to the two examples used in this work, there are several other binary systems with high Shapiro delay, such as MSP J1909-3744 \cite{Shamohammadi2023}, MSP J0740+6620 \cite{Cromartie2020}, MSP J1012-4235 \cite{Gautam2024}.
However, none of these sources turned out to be satisfactory.
If one wants to probe the unexcluded axion parameter space (e.g. $\sim10^{-12}\,{\rm GeV^{-1}}$) with PSR J1614-2230, the minimum distance between the WD and the line of sight direction must be less than $d \lesssim 4.2 \times 10^{7} \rm{m}$, implying an orbital inclination of $i \sim 89.85^{\circ}$, which would be a very rare binary system with high inclination angle. 
We hope that future observations can lead to the discovery of more suitable binaries for further probing of axion parameter space.

\begin{acknowledgments}
This work is supported by the National Key Research and Development Program of China (Grant No. 2022YFF0503304) and the Guangxi Talent Program (“Highland of Innovation Talents”).
\end{acknowledgments}

\bibliography{apssamp}

\end{document}